\documentclass[preprint]{aastex}
\usepackage{rotating} 
\usepackage{amsmath,amssymb}
\usepackage{subfigure}
\usepackage{bm}

\begin{document}

\title{Non coherent continuum scattering as a line polarization mechanism}

\author{T. del Pino Alem\'an\altaffilmark{1,2}, R. Manso Sainz\altaffilmark{1,2} and J. Trujillo Bueno\altaffilmark{1,2,3}} \altaffiltext{1}{Instituto de Astrof\'{\i}sica de Canarias, 38205, La Laguna, Tenerife, Spain}\altaffiltext{2}{Departamento de Astrof\'\i sica, Facultad de F\'\i sica, Universidad de La Laguna, Tenerife, Spain}\altaffiltext{3}{Consejo Superior de Investigaciones Cient\'{\i}ficas, Spain} \email{tanausu@iac.es, rsainz@iac.es, jtb@iac.es}

\begin{abstract}
Line scattering polarization can be strongly affected by Rayleigh scattering by neutral hydrogen and Thompson scattering by free electrons. Often a continuum depolarization results, but the Doppler redistribution produced by the continuum scatterers, which are light (hence, fast), induces more complex interactions between the polarization in spectral lines and in the continuum. Here we formulate and solve the radiative transfer problem of scattering line polarization with non-coherent continumm scattering consistently. The problem is formulated within the spherical tensor representation of atomic and light polarization. The numerical method of solution is a generalization of the Accelerated Lambda Iteration that is applied to both, the atomic system and the radiation field. We show that the redistribution of the spectral line radiation due to the non coherence of the continuum scattering may modify significantly the shape of the emergent fractional linear polarization patterns, even yielding polarization signals above the continuum level in intrinsically unpolarizable lines.
\end{abstract}

\keywords{Polarization - radiative transfer - scattering - stars: atmospheres - Sun: atmosphere}

\section{Introduction}\label{S1}

The solar spectrum observed close to the solar limb is linearly polarized. The polarization of the continuum, first observed by \cite{Lyot1948}, is mainly produced by scattering at neutral hydrogen (Rayleigh scattering) and free electrons (Thomson scattering). But spectral lines, linearly polarized by scattering processes, show incredibly rich and complex polarizations patterns (e.g., \citealt{StenfloTwerenboldHarvey1983,bStenfloTwerenboldHarvey1983,StenfloKellerGandorfer2000,BGandorfer2000,BGandorfer2002,BGandorfer2005}). This {\em second solar spectrum} (\citealt{Ivanov1991,StenfloKeller1997}) has been the subject of many theoretical investigations because of its diagnostic potential for the magnetism (and thermodynamics) of the solar atmosphere; but some polarization patterns are not yet well understood (see \citealt{Trujillo2009} for a review).

Scattering line polarization is usually modeled independently of the continuum polarization. The continuum polarization is modeled as a coherent scattering process (\citealt{Debarbat1970,FluriStenflo1999,TrujilloShchukina2009}), which is a suitable approximation far from spectral lines. Thomson and Rayleigh scattering are coherent in the scatterer's frame (e.g., \citealt{BChandrasekhar1950}). However, the Doppler broadening corresponding to the thermal velocity of electrons and hydrogen atoms is several times the width of most spectral lines, which may lead to redistribution between the polarization of the spectral line and the nearby continuum (e.g., \citealt{BLandiLandolfi2004}, henceforth LL04). 

The effect of non-coherent continuum scattering in radiative transfer was considered by \cite{Munch1948}, but his work did not include light polarization. He showed that the effect of the non-coherence on the intensity line spectrum is to broaden the profile and to make it shallower. The treatment of the Rayleigh and Thomson scattering was first extended to the non-coherent and polarized case by \cite{SenLee1961}, who studied some of the effects that this phenomenon may have on the emergent spectral line radiation. These initial steps were later continued by other researchers (e.g., \citealt{NagendraRangarajanRao1993,Rangarajan1999}) who studied the problem of non-coherent electron scattering and partial frequency redistribution on the polarization of resonance lines, pointing out the significance of electron scattering redistribution in the far wings of the line polarization profile. This result has been recently confirmed by \cite{SupriyaNagendraSampoornaRavindra2012} after solving the same type of problem through the application of more efficient numerical radiative transfer methods.

In this paper we treat the radiation transfer problem of resonance line polarization taking into account its interaction with non-coherent scattering in the continuum. We treat the Rayleigh and Thomson redistribution as angle independent (angle averaged redistribution), and the line emission and absorption using the two-level atom model with unpolarized lower level in the limit of complete frequency redistribution (CRD). To solve the relevant equations, formulated within the framework of the density matrix theory (see LL04), we develop an efficient Jacobian iterative method, which can be considered as a generalization of that proposed by \cite{TrujilloManso1999} for the CRD line transfer case. We apply this numerical method to solve the radiation transfer problem in a Milne-Eddington atmosphere and in a stratified model atmosphere with a temperature minimum and a chromospheric temperature rise. We study the effects of the non-coherence of the continuum scattering on intrinsically unpolarizable (transition between upper and lower levels with angular momentum $J_{u} = J_{\ell} = 1/2$) and polarizable ($J_{u}=1$ and $J_{\ell}=0$) lines. In particular, we show the possibility of generating ``emission" fractional linear polarization features (i.e., with larger polarization than in the adjacent continuum) in the core of intrinsically unpolarizable spectral lines.

\section{Formulation of the problem}

\begin{figure}[tb]
\centering
\includegraphics[width=8.6cm]{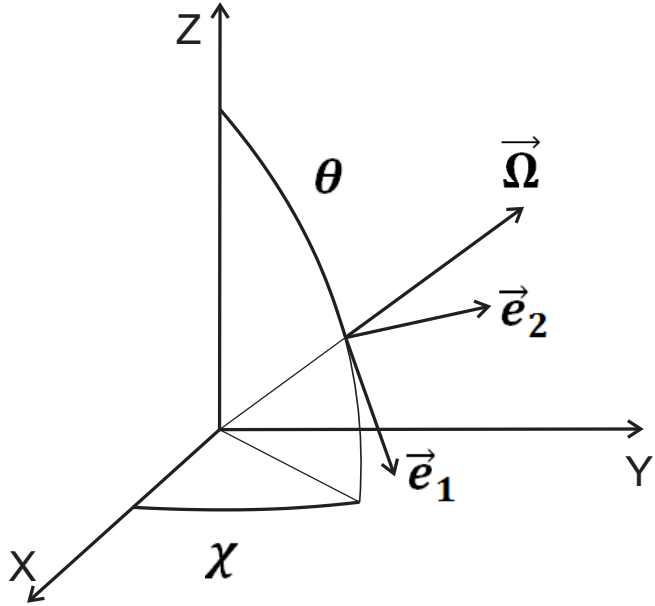}
 \caption{Reference system for polarization. $\theta$ and $\chi$ are the polar and the azimutal angles of the ray under consideration, respectively. $\vec{\Omega}$ is the propagation direction, $\vec{e_{1}}$ is perpendicular to $\vec{\Omega}$ and is on the meridian plane, and $\vec{e_{2}}$ is perpendicular to $\vec{\Omega}$ and $\vec{e_{1}}$. In all the equations, the direction of positive Stokes $Q$ is taken along $\vec{e_{1}}$, i.e., perpendicular to the projected limb.}
\label{FIGAxis}
\end{figure}

    We consider resonance line polarization (assuming the CRD and two-level atom model without stimulated emission) in the presence of a polarized continuum in a plane-parallel, static and non-magnetic atmosphere. Due to the symmetry of the problem, the radiation field is rotationally invariant with respect to the vertical direction (which we choose to be the $z$ axis) and it is thus linearly polarized along a direction either parallel or perpendicular to the projected limb. Using the reference system for polarization of Fig. \ref{FIGAxis} the radiation field is characterized by just the Stokes parameters $I$ and $Q$. Assuming that the lower level of the transition is unpolarized (either its total angular momentum is $J_{\ell} = 0$ or $1/2$, or collisions dominate its excitation), the absorption process is isotropic and the radiative transfer equations for $I$ and $Q$ 
at frequency $\nu$ and propagation direction $\vec{\Omega}$ are
\begin{subequations}\label{EQRT}\begin{align}
  \frac{d I}{d \tau} = & I - S_{I} , \label{EQRTI} \\ 
  \frac{d Q}{d \tau} = & Q - S_{Q} , \label{EQRTQ}
\end{align}\end{subequations}
where $d\tau = -\chi d\ell$ is the element of optical distance (where $\ell$ is the geometrical distance measured along the ray direction), $\chi = \chi_{l}\phi\left(x\right) + \chi_{c}$ is the total absorption coefficient; $\chi_{l}$ and $\chi_{c} = \kappa + \sigma$ are the integrated line and total continuum absorption coefficients, respectively; $\kappa$ and $\sigma$ are the thermal and scattering continuum absorption coefficients; $\phi\left(x\right)$ is the line absorption profile, and $x=\left(\nu - \nu_{0}\right)/\Delta\nu_{D}$ is the frequency separation from the resonance frequency $\nu_{0}$ in units of the Doppler width $\Delta\nu_{D}$. $S_{I}$ and $S_{Q}$ are the source functions, which for a two-level atom with polarized continuum are 
\begin{subequations}\label{EQS}\begin{align}
S_{I}\left(x\right)  = &  r_{x} S^{l}_{I} + \left(1 - r_{x} \right) S_{I}^{c}\left(x\right) 
, \label{EQSI} \\
S_{Q}\left(x\right) = & r_{x} S^{l}_{Q} + \left(1 - r_{x} \right) S_{Q}^{c}\left(x\right) 
, \label{EQSQ}
\end{align}\end{subequations}
where $r_{x} = \chi_{l} \phi\left(x\right) / \left(\chi_{l} \phi\left(x\right) + \chi_{c}\right)$.
The line source functions are expressed in terms of the excitation state of the upper level of the transition. In this case, due to symmetry, the only non-zero spherical components are $\rho^{0}_{0}$ ($\sqrt{2J_{u} +1}$ times the total population) and $\rho^{2}_{0}$ (alignment coefficient) of the density matrix (\citealt{BBlum1981}) of the upper level, and the line source functions are (e.g., \citealt{TrujilloManso1999})
\begin{subequations}\label{EQSline}\begin{align}
S_{I}^{l} = & \frac{2 h \nu^{3}}{c^{2}}\frac{2 J_{\ell} + 1}{\sqrt{2 J_{u} + 1}} \left[\rho^{0}_{0} + \frac{w^{\left(2\right)}_{J_{u} J_{\ell}}}{2 \sqrt{2}}\left(3 \mu^{2} - 1\right) \rho^{2}_{0}\right]
, \label{EQSIline} \\
S_{Q}^{l} = & \frac{2 h \nu^{3}}{c^{2}}\frac{2 J_{\ell} + 1}{\sqrt{2 J_{u} + 1}} \frac{3 w^{\left(2\right)}_{J_{u} J_{\ell}}}{2 \sqrt{2}}\left(\mu^{2} - 1\right) \rho^{2}_{0} 
, \label{EQSQline}
\end{align}\end{subequations}
where $w^{(2)}_{J_{u}J_{\ell}}$ is a numerical coefficient which depends on the total angular momentum of the levels involved in the transition (Table 10.1 in LL04; e.g., $w^{\left(2\right)}_{1 0} = 1$, $w^{\left(2\right)}_{\frac{1}{2} \frac{1}{2}} = 0$). $\mu = \cos{\theta}$, where $\theta$ is the angle of the line of sight (LOS) to $z$ (see Fig. \ref{FIGAxis}). 

The density matrix elements are obtained from the following statistical equilibrium equations (\citealt{TrujilloManso1999}):
\begin{subequations}\label{EQrho}\begin{align}
\frac{2 h \nu^{3}}{c^{2}}\frac{2 J_{\ell} + 1}{\sqrt{2 J_{u} + 1}}\rho^{0}_{0} = & \left(1 - \epsilon\right)\bar{J}_{0}^{0} + \epsilon B_{\nu}
, \label{EQrho00} \\
\frac{2 h \nu^{3}}{c^{2}}\frac{2 J_{\ell} + 1}{\sqrt{2 J_{u} + 1}}\rho^{2}_{0} = & \frac{1-\epsilon}{1 + \left(1-\epsilon\right)\delta^{\left(2\right)}}w^{\left(2\right)}_{J_{u}J_{\ell}}\bar{J}_{0}^{2}
, \label{EQrho20}
\end{align}\end{subequations}
where $B_{\nu}$ is the Planck function, $\epsilon = C_{u\ell}/(A_{u\ell}+C_{u\ell})$ is the collisional destruction probability due to inelastic collisions ($C_{u\ell}$ and $A_{u\ell}$ are the collisional de-excitation rate and Einstein coefficient for spontaneous emission, respectively) and $\delta^{\left(2\right)}=D^{\left(2\right)}/A_{u\ell}$ ($D^{\left(2\right)}$ is the depolarizing rate of the level due to elastic collisions with neutral hydrogen). 

The radiation field tensors in Eqs. \eqref{EQrho} are given by
\begin{subequations}\label{EQJbar}\begin{align}
\bar{J}^{0}_{0} = & \int dx \, \phi\left(x\right) J^{0}_{0}\left(x\right)
,\label{EQJ00bar} \\
\bar{J}^{2}_{0} = & \int dx \, \phi\left(x\right) J^{2}_{0}\left(x\right)
, \label{EQJ20bar}\end{align}\end{subequations} 
where $J^{0}_{0}\left(x\right)$ and $J^{2}_{0}\left(x\right)$ are the frequency-dependent radiation field tensors defined as (LL04)
\begin{subequations}\label{EQJx}\begin{align} 
J^{0}_{0}\left(x\right)  = & \frac{1}{2}\int_{-1}^{1} d \mu' I\left(x, \mu'\right) 
,\label{EQJ00x} \\ \begin{split} 
J^{2}_{0}\left(x\right)  = & \frac{1}{4 \sqrt{2}}\int_{-1}^{1} d \mu' \Big[\left(3 \mu'^{2} - 1\right) I\left(x, \mu'\right)  \\  & + 3 \left(\mu'^{2} - 1\right) Q\left(x, \mu'\right)\Big]
.\label{EQJ20x} 
\end{split}\end{align}\end{subequations}

The source functions for the background continuum in Eqs. \eqref{EQS}, taking into account thermal emission and scattering, can be expressed as (e.g., \citealt{MansoTrujillo2011})
\begin{subequations}\label{EQScont}\begin{align}
 S_{I}^{c}\left(x\right) = &  s B_{\nu} + \left(1 - s\right)\left[\breve{J}_{0}^{0}\left(x\right) + \frac{1}{2 \sqrt{2}} \left(3 \mu^{2} - 1\right) \breve{J}^{2}_{0}\left(x\right)\right]
, \label{EQSIcont} \\
 S_{Q}^{c}\left(x\right) = & \left(1 - s\right)\frac{3}{2 \sqrt{2}} \left(\mu^{2} - 1\right) \breve{J}^{2}_{0}\left(x\right)
, \label{EQSQcont}
\end{align}\end{subequations}
  where $s = \kappa / \chi_{c}$, with the convolved radiation field tensors 
\begin{subequations}\label{EQJconv}\begin{align}
\breve{J}^{0}_{0}\left(x\right) = & \int d x' \phi_{c}\left(x,x'\right) J^{0}_{0}\left(x'\right)
, \label{EQJ00conv} \\
\breve{J}^{2}_{0}\left(x\right) = & \int d x' \phi_{c}\left(x,x'\right) J^{2}_{0}\left(x'\right)
, \label{EQJ20conv}
\end{align}\end{subequations}
where $x'$ and $x$ are the frequencies of the incident and scattered photons, respectively.  The convolution profile $\phi_{c}\left(x,x'\right)$ accounts for the frequency redistribution caused by the Doppler effect, due to the velocity distribution of the scatterers (electrons for Thomson scattering; hydrogen and helium for Rayleigh scattering).

Thomson scattering is coherent in the scatterer's reference system. We take into account the Doppler shifts due to the motions of the electrons relative to the laboratory frame by averaging over their velocity distribution, which we assume to be Maxwellian. We also take the average over the solid angle (greatly reducing the computational cost) because the angular distribution is less important than the frequency distribution (\citealt{BMihalas1978}) and the difference with the angle-dependent distribution function is small for optically thick atmospheres (\citealt{SupriyaNagendraSampoornaRavindra2012}). The final expression for the angle averaged convolution profile is (\citealt{Hummer1962,BMihalas1978})
\begin{equation}
\phi_{c}\left(x,x'\right)=\phi_{c}\left(y\right) = \frac{1}{w}\left[\frac{e^{-y^{2}}}{\sqrt{\pi}} - y\cdot {\rm erfc}\left(y\right)\right]
,\label{EQconvprof}
\end{equation}
with
\begin{equation}
y = \left|\frac{x - x'}{2 w}\right|
,\label{EQy}
\end{equation}
where $w$ is the ratio between the Doppler widths of the perturbers and the atom of interest. 

Rayleigh scattering is produced in the far wings of the Lyman lines of neutral hydrogen and helium. We may consider that the scattering in the very far wings of a resonance line is essentially coherent in the scatterer rest frame (e.g., \citealt{BMihalas1978}) and the above discussion for Thomson scattering applies also to Rayleigh scattering taking into account the different value of $w$. 

If we consider the simultaneous contribution of Thomson and Rayleigh scattering, $\sigma = \sigma_{\rm{T}} + \sigma_{\rm{R}}$, different source function terms appear for each convolution kernel (Thomson and Rayleigh) and convolved radiation field tensor. For simplicity, we will not write explicitly such expressions here. To avoid a lengthy expression, we consider explicitly only one of the contributions of the background continuum scattering; accounting for additional contributions is straightforward.

The source functions in Eqs. \eqref{EQS} may be expressed in a more simple and symmetric form as
\begin{subequations}\label{EQSt}\begin{align}
S_{I}\left(x\right) = & S^{0}_{0}\left(x\right) + \frac{1}{2 \sqrt{2}}\left(3\mu^{2} - 1\right) S^{2}_{0}\left(x\right) 
, \label{EQSIt} \\
S_{Q}\left(x\right) = & \frac{3}{2 \sqrt{2}} \left(\mu^{2} - 1\right) S^{2}_{0}\left(x\right)
. \label{EQSQt}
\end{align}\end{subequations}
Here $S^{K}_{Q}\left(x\right)$ are the frequency-dependent source function tensors:
\begin{subequations}\label{EQSKQdef}\begin{align}
S^{0}_{0}\left(x\right) = & r_{x}S^{0}_{0} + \left(1 - r_{x}\right)S^{0}_{0}{}^{c}\left(x\right)
,\label{EQS00def} \\
S^{2}_{0}\left(x\right) = & r_{x}w^{\left(2\right)}_{J_{u}J_{\ell}} S^{2}_{0} + \left(1 - r_{x}\right)S^{2}_{0}{}^{c}\left(x\right)
,\label{EQS20def}
\end{align}\end{subequations}
where the $S^{K}_{0} = (2 h \nu_{0}^3/c^{2})(2J_{\ell} + 1)/\sqrt{(2J_{u} +1)}\rho^{K}_{0}$ tensors are given by
\begin{subequations}\label{EQSKQldef}\begin{align}
S^{0}_{0} = & \left(1 - \epsilon\right)\bar{J}_{0}^{0} + \epsilon B_{\nu}
,\label{EQS00ldef} \\
S^{2}_{0} = & \frac{1-\epsilon}{1 + \left(1-\epsilon\right)\delta^{\left(2\right)}}w^{\left(2\right)}_{J_{u}J_{\ell}}\bar{J}_{0}^{2}
,\label{EQS20ldef}
\end{align}\end{subequations} 
and the continuum frequency-dependent tensors by
\begin{subequations}\label{EQSKQcdef}\begin{align}
S^{0}_{0}{}^{c}\left(x\right) = & \left(1 - s\right)\breve{J}_{0}^{0}\left(x\right) + s B_{\nu}
,\label{EQS00cdef} \\
S^{2}_{0}{}^{c}\left(x\right) = & \left(1 - s\right)\breve{J}_{0}^{2}\left(x\right)
.\label{EQS20cdef}
\end{align}\end{subequations}

\section{Numerical Method of Solution}

   Equations \eqref{EQRT} together with Eqs.\eqref{EQS}-\eqref{EQJconv} or, equivalently, \eqref{EQSt}-\eqref{EQSKQcdef}, form a coupled system of integro-differential equations which we solve numerically. We consider an iterative method of solution: if an estimate of the source functions is given, Eqs.\eqref{EQRT} can be integrated for a given set of boundary conditions; from the radiation field thus calculated we reevaluate the $S^{K}_{0}$ and $S^{K}_{0}\left(x\right)$ tensors which are in turn used to recalculate the new source functions and hence a new radiation field estimate. The formal solution integration (Sect. \ref{S31}) is based on the short-characteristics (SC) method (\citealt{KunaszAuer1988}); in order to guarantee convergence, the iterative scheme  (Sect. \ref{S32}) is a generalization of the Accelerated Lambda Iteration (\citealt{OlsonAuerBuchler1986}) developed by \cite{TrujilloManso1999}, which is based on the Jacobi method.

\subsection{Formal Solution}\label{S31}

  If the source functions are given, Eqs.\eqref{EQRT} can be integrated explicitly between two spatial points $i$ and $j$, for a given frequency and angle:
\begin{equation}
I\left(x;j\right) = I\left(x;i\right) e^{-\Delta\tau_{ij}} + \int_{0}^{\Delta\tau_{ij}} S\left(x;t\right)e^{- t} dt 
, \label{EQI}
\end{equation}
  and analogously for $Q$.
  In Eq.\eqref{EQI}, $\Delta\tau_{ij}$ is the optical distance along the ray between points \emph{i} and \emph{j} at the reduced frequency $x$.

We assume that the source function varies parabolically between three consecutive points M, O and P:  O is the point where we want to calculate the Stokes parameters, while M and P are respectively the preceding and following points according to the propagation direction. Eq.\eqref{EQI} can then be rewritten as (\citealt{KunaszAuer1988})
\begin{equation}\begin{split}
I\left(\rm{O}\right) = & I\left(\rm{M}\right) e^{-\Delta\tau_{\rm{M}}} + \Psi_{\rm{M}}\left(\rm{O}\right)S_{I}\left(\rm{M}\right) \\ & + \Psi_{\rm{O}}\left(\rm{O}\right)S_{I}\left(\rm{O}\right)  + \Psi_{\rm{P}}\left(\rm{O}\right)S_{I}\left(\rm{P}\right)
, \label{EQISC}
\end{split}\end{equation}
where $I(\rm{O})$ and $I(\rm{M})$ are the intensities at points O and M, $\Delta\tau_{\rm{M}}$ is the optical distance between points M and O; $S_{I}(\rm{M})$, $S_{I}(\rm{O})$ and $S_{I}(\rm{P})$ are the values of the intensity source function at the points M, O and P, respectively, and $\Psi_{\rm{M}}(\rm{O})$, $\Psi_{\rm{O}}(\rm{O})$ and $\Psi_{\rm{P}}(\rm{O})$ are three functions that only depend on the optical distance between the local point (O in this case) and the preceding and following points (M and P in this equation).

\subsection{Iterative Scheme}\label{S32}

   Equation \eqref{EQISC} expresses the intensity at point O as a linear combination of the source function at adjacent points in the atmosphere and the intensity at a {\em previous} point M along the ray. The same scheme can in turn be applied to the {\em previous} point and repeated all the way back to the boundary where the incoming radiation is given. Therefore, the Stokes parameters at a point $i$ along a given ray in the atmosphere can be expressed as
\begin{subequations}\label{EQRTLD}\begin{align}
I\left(x,\mu;i\right) = \sum_{j = 1}^{N_{z}} \Lambda\left(x,\mu;i,j\right)S_{I}\left(x,\mu;j\right) + T_{I}\left(x,\mu;i\right)
, \label{EQRTLID} \\
Q\left(x,\mu;i\right) = \sum_{j = 1}^{N_{z}} \Lambda\left(x,\mu;i,j\right)S_{Q}\left(x,\mu;j\right) + T_{Q}\left(x,\mu;i\right)
, \label{EQRTLQD}
\end{align}\end{subequations}
where the $\Lambda\left(x,\mu;i,j\right)$ coefficients depend on the optical distances between points ``$i$'' and ``$j$'', 
$T_{I,Q}\left(x,\mu;i\right)$ are the {\em transmitted} Stokes parameters from the boundary, and $N_{z}$ the number of spatial grid points.
Averaging these expressions over the angles (Eqs. \eqref{EQJx}) and taking into account the dependence of the source function components 
on $S^{K}_{0}$ (Eqs. \ref{EQSt}), the radiation field tensors at a point $i$ in the atmosphere can be expressed as
\begin{subequations}\label{EQJxnum}\begin{align}\begin{split}
J^{0}_{0}\left(x;i\right) = & \sum_{j=1}^{N_{z}}\Lambda^{0}_{0}\left(x;i,j\right)S^{0}_{0}\left(x;j\right) \\ & + \sum_{j = 1}^{N_{z}}\Lambda^{0}_{2}\left(x;i,j\right)S^{2}_{0}\left(x;j\right) + T^{0}_{0}\left(x;i\right)
, \label{EQJ00xnum}\end{split} \\ \begin{split}
J^{2}_{0}\left(x;i\right) = & \sum_{j = 1}^{N_{z}}\Lambda^{2}_{0}\left(x;i,j\right)S^{0}_{0}\left(x;j\right) \\ & + \sum_{j = 1}^{N_{z}}\Lambda^{2}_{2}\left(x;i,j\right)S^{2}_{0}\left(x;j\right) + T^{2}_{0}\left(x;i\right)
.\label{EQJ20xnum}
\end{split}\end{align}\end{subequations}
The explicit expressions for $\Lambda^{K}_{K'}\left(x;i,j\right)$ and $T^{K}_{0}\left(x;i\right)$ in terms of $\Lambda\left(x,\mu;i,j\right)$ and $T_{I,Q}\left(x;i\right)$ are given in the Appendix.
It is important to emphasize that we do not need to calculate them explicity (except for the diagonal elements); they are implicitly evaluated according to the SC algorithm described in the previous section. Equations \eqref{EQJxnum} are only convenient to derive the iterative scheme, as we will now show.

Let $S^{0}_{0}{}^{\rm old}$, $S^{2}_{0}{}^{\rm old}$, $J^{0}_{0}{}^{\rm old}\left(x\right)$ and $J^{2}_{0}{}^{\rm old}\left(x\right)$ be estimates at some iterative step of the atomic and radiation field tensors, and $S^{0}_{0}{}^{\rm old}\left(x\right)$ and $S^{2}_{0}{}^{\rm old}\left(x\right)$ the corresponding frequency dependent source function tensors derived from them using Eqs. \eqref{EQSKQdef}-\eqref{EQSKQcdef}. Let $J^{0}_{0}{}^{\dagger}\left(x\right)$ and $J^{2}_{0}{}^{\dagger}\left(x\right)$ be the values of the radiation field tensors obtained through the formal solution of the radiative transfer equation (Sect. \ref{S31}) using the above-mentioned ``old" quantities ---formally, using $S^{K}_{0}{}^{\rm old}$ on the right hand side of Eqs. \eqref{EQJxnum}.

If we used $J^{K}_{0}{}^{\dagger}\left(x\right)$ to calculate the corresponding $\bar{J}^{K}_{0}{}$ and $\breve{J}^{K}_{0}\left(x\right)$ (Eqs. \eqref{EQJbar} and \eqref{EQJconv}, respectively), and then, Eqs. \eqref{EQSKQdef}-\eqref{EQSKQcdef} to obtain new estimates of $S^{K}_{0}$ and $S^{K}_{0}\left(x\right)$, we would have a generalization of the Lambda iteration scheme which is known to have very poor convergence properties (e.g., \citealt{BMihalas1978}).

In order to improve the convergence rate, let's consider Eqs. \eqref{EQJxnum}. Formally, now we shall calculate the radiation field tensors at a given point ``$i$" from the $S^{K}_{0}{}^{\rm old}\left(x; j\right)$ at all grid points $j\ne i$, and the yet unknown ``new" value $S^{K}_{0}\left(x; i\right)$ at point ``$i$". Rearranging terms:

\begin{subequations}\label{EQJxali}\begin{align}\begin{split}
J^{0}_{0}\left(x;i\right) \approx &  J^{0}_{0}{}^{\dagger}\left(x;i\right) + \Lambda^{0}_{0}\left(x;i,i\right)\delta S^{0}_{0}\left(x;i\right) \\ & + \Lambda^{0}_{2}\left(x;i,i\right)\delta S^{2}_{0}\left(x;i\right)
,\label{EQJ00xali}\end{split} \\ \begin{split}
J^{2}_{0}\left(x;i\right) \approx & J^{2}_{0}{}^{\dagger}\left(x;i\right) + \Lambda^{2}_{0}\left(x;i,i\right)\delta S^{0}_{0}\left(x;i\right) \\  & + \Lambda^{2}_{2}\left(x;i,i\right)\delta S^{2}_{0}\left(x;i\right)
,\label{EQJ20xali}
\end{split}\end{align}\end{subequations}
where
\begin{equation}
\delta S^{K}_{0}\left(x;i\right) = S^{K}_{0}\left(x;i\right) - S^{K}_{0}{}^{\rm{old}}\left(x;i\right)
. \label{EQSKQdelta}
\end{equation}

Equations \eqref{EQJxali} show how to actually compute these new radiation field tensors: $J^{K}_{0}{}^\dagger(x)$ is calculated exactly as explained in the previous paragraph; the diagonal components of the operators $\Lambda^K_{K'}(x; i.i)$ can be efficiently computed while performing the formal solution (see \citealt{AsensioTrujillo2006}); finally, the yet-to-be-obtained $S^{K}_{0}\left(x\right)$ elements are kept explicitly; the whole iterative scheme will be obtained from consistently applying these expressions for $J^{K}_{0}\left(x\right)$ and finally solving the resulting system of algebraic equations for $S^{K}_{0}\left(x\right)$. It can be demonstrated that in solar-like atmospheres the convergence rate of this iterative scheme is practically unaffected if one retains only the zeroth-order Lambda operator $\Lambda^{0}_{0}$ while putting $\Lambda^{0}_{2}=\Lambda^{2}_{0}=\Lambda^{2}_{2}=0$ in Eqs. \eqref{EQJxali} (see \citealt{TrujilloManso1999}). Therefore, we shall develop this simplified Jacobian iterative scheme in the following.

Using Eq. \eqref{EQJ00xali} we calculate the average over the line profile of the radiation field tensor:
\begin{equation}
\bar{J}^{0}_{0}\left(i\right) = \bar{J}^{0}_{0}{}^{\dagger}\left(i\right) + \int dx' \phi\left(x';i\right)\Lambda^{0}_{0}\left(x';i,i\right)\delta S^{0}_{0}\left(x';i\right)
. \label{EQMT03}
\end{equation}
Substituting this equation for the mean radiation field tensor into Eq. \eqref{EQS00ldef} for the source function $S^{0}_{0}$ and subtracting $S^{0}_{0}{}^{\rm old}\left(i\right)$, we find

\begin{equation}\begin{split}
\delta S^{0}_{0}\left(i\right) = & \left(1 - \epsilon\right)\Big[\bar{J}^{0}_{0}{}^{\dagger}\left(i\right) - \bar{J}^{0}_{0}{}^{\rm old}\left(i\right) \\ & + \int dx' \phi\left(x';i\right) \Lambda^{0}_{0}\left(x';i,i\right) \delta S^{0}_{0}\left(x';i\right)\Big]
, \label{EQMT4}
\end{split}\end{equation}
where we have made explicit the height dependence of $\phi\left(x\right)$ (the dependence of $\epsilon$, $r_{x}$ and $s$ is kept implicit). If we substitute into this equation the expression of the source function $S^{0}_{0}\left(x\right)$ of Eq. \eqref{EQS00def}, we obtain
\begin{equation}\begin{split}
\delta S^{0}_{0}\left(i\right) = & \left(1 - \epsilon\right)\Big[\bar{J}^{0}_{0}{}^{\dagger}\left(i\right) - \bar{J}^{0}_{0}{}^{\rm old}\left(i\right)  \\ & + \int dx' r_{x'} \phi\left(x';i\right) \Lambda^{0}_{0}\left(x';i,i\right) \delta S^{0}_{0}\left(i\right) \\ & + \int dx' \left(1 - r_{x'}\right) \phi\left(x';i\right) \Lambda^{0}_{0}\left(x';i,i\right) \delta S_{0}^{0}{}^{c}\left(x';i\right)\Big]
.  \label{EQMT5}
\end{split}\end{equation}
Moreover, defining
\begin{equation}
\bar{\Lambda}^{0}_{0}\left(i,i\right) = \int dx' \phi\left(x';i\right)r_{x'}\Lambda^{0}_{0}\left(x';i,i\right)
,\label{EQL00bar}
\end{equation}
taking into account that
\begin{equation}
\delta S^{0}_{0}{}^{c}\left(x;i\right) = \left(1 - s\right)\int dx' \phi_{c}\left(x,x';i\right)\delta J^{0}_{0}\left(x';i\right)
, \label{EQMT04}
\end{equation}
where
\begin{equation}
\delta J^{0}_{0}\left(x;i\right) = J^{0}_{0}\left(x;i\right) - J^{0}_{0}{}^{\rm old}\left(x;i\right)
, \label{EQMT05}
\end{equation}
noting also that
\begin{equation}\begin{split}
&\int dx' \left(1 - r_{x'}\right)\phi\left(x';i\right)\Lambda^{0}_{0}\left(x';i,i\right) \\ 
&\times\int dx'' \phi_{c}\left(x',x'';i\right)\delta J^{0}_{0}\left(x'';i\right) = \\
&\int dx' \delta J^{0}_{0}\left(x';i\right)  \\ 
&\times\int dx'' \phi\left(x'';i\right)\left(1 - r_{x''}\right)\Lambda^{0}_{0}\left(x'';i,i\right)\phi_{c}\left(x',x'';i\right)
, \label{EQMT06}
\end{split}\end{equation}
and using Eqs.\eqref{EQS00cdef} and \eqref{EQJ00conv}, we find that the correction to the line source function is
\begin{equation}\begin{split}
\delta S^{0}_{0}\left(i\right)  = & \frac{\left(1 - \epsilon\right)}{1 - \left(1 - \epsilon\right) \bar{\Lambda}^{0}_{0}\left(i,i\right)}\Big[ \bar{J}^{0}_{0}{}^{\dagger}\left(i\right) - \bar{J}^{0}_{0}{}^{\rm old}\left(i\right) \\ & + \left(1 - s\right)\int dx' \delta J^{0}_{0}\left(x';i\right) \\ & \times\int dx'' \phi\left(x'';i\right)\left(1 - r_{x''}\right)\Lambda^{0}_{0}\left(x'';i,i\right)\phi_{c}\left(x',x'';i\right)\Big]
, \label{EQMT7}
\end{split}\end{equation}
  Applying the same reasoning to the continuum source function, from Eq.\eqref{EQJ00xali}, with $\Lambda_{0}^{2} = 0$
\begin{equation}
J^{0}_{0}\left(x;i\right) = J^{0}_{0}{}^{\dagger}\left(x;i\right) + \Lambda^{0}_{0}\left(x;i,i\right)\delta S^{0}_{0}\left(x;i\right)
. \label{EQMT9}
\end{equation}
  Taking the variation of the field tensor,
\begin{equation}\begin{split}
\delta J^{0}_{0}\left(x;i\right) = & J^{0}_{0}{}^{\dagger}\left(x;i\right) - J^{0}_{0}{}^{\rm old}\left(x;i\right) \\ & + r_{x} \Lambda^{0}_{0}\left(x;i,i\right) \delta S^{0}_{0}\left(i\right) \\ & + \left(1 - r_{x}\right) \left(1 - s\right) \Lambda^{0}_{0}\left(x;i,i\right) \\ & \times\int dx'\phi_{c}\left(x,x';i\right)\delta J^{0}_{0}\left(x';i\right)
, \label{EQMT10}
\end{split}\end{equation} 
  and substituting Eq.\eqref{EQMT7} into Eq. \eqref{EQMT10}, after gathering the terms in $\delta J^{0}_{0}\left(x';i\right)$, we obtain:
\begin{equation}\begin{split}
 & \int dx'\Big[ \delta\left(x-x'\right) - \left(1 - s\right)\left(1 - r_{x}\right) \phi_{c}\left(x,x';i\right)\Lambda_{0}^{0}\left(x;i,i\right) \\ & - \frac{ \Lambda^{0}_{0}\left(x;i,i\right) r_{x} \left(1 - \epsilon\right)\left(1 - s\right)}{1 - \left(1 - \epsilon\right)\bar{\Lambda}^{0}_{0}\left(i,i\right)}\\ & \times\int dx'' \phi\left(x'';i\right)\phi_{c}\left(x',x'';i\right) \left(1 - r_{x''}\right) \Lambda^{0}_{0}\left(x'';i,i\right)\Big] \\ & \times\delta J^{0}_{0}\left(x';i\right) = J^{0}_{0}{}^{\dagger}\left(x;i\right) - J^{0}_{0}{}^{\rm old}\left(x;i\right) \\ & + \frac{\Lambda^{0}_{0}\left(x;i,i\right) r_{x} \left(1 - \epsilon\right)\left(\bar{J}^{0}_{0}{}^{\dagger}\left(i\right) - \bar{J}^{0}_{0}{}^{\rm old}\left(i\right)\right)}{1 - \left(1 - \epsilon\right) \bar{\Lambda}^{0}_{0}\left(i,i\right)}
. \label{EQMT11}
\end{split}\end{equation}
  The discretization of Eq. \eqref{EQMT11} in the frequency domain gives a linear system of $N_{x}$ equations (with $N_{x}$ the number of frequency points) for $\delta J^{0}_{0}\left(x\right)$. Substitution of this solution into Eq. \eqref{EQMT7} completes the iterative scheme for $S^{0}_{0}\left(x\right)$.

  As pointed out by \cite{TrujilloManso1999} and stated above, the solution of standard resonance line polarization problems using methods based on Jacobi iteration can simply rely on the diagonal of the $\Lambda^{0}_{0}$ operator. The resulting equation for $\delta S^{2}_{0}\left(x\right)$ is thus formally equivalent to consider Lambda iteration for $S_{Q}$. However, it is crucial to note that $J^{2}_{0}\left(x\right)$ is improved at the rate of $\delta J^{0}_{0}\left(x\right)$, because the anisotropy tensor $J^{2}_{0}\left(x\right)$ is dominated by the Stokes $I$ parameter which is, in turn, basically set by the values of $S^{0}_{0}\left(x\right)$. From Eqs.\eqref{EQSQline}, \eqref{EQSQcont} and \eqref{EQrho20}
\begin{equation}\begin{split}
\delta S^{2}_{0}\left(x;i\right) = & r_{x}\frac{3w_{J_{u}J_{\ell}}^{\left(2\right)}}{2\sqrt{2}}\left(\mu^{2}-1\right)\frac{1 - \epsilon}{1 + \left(1 - \epsilon\right)\delta^{\left(2\right)}}\bar{J}^{2}_{0}{}^{\dagger}\left(i\right) \\ & + \left(1-r_{x}\right)\left(1-s\right)\frac{3}{2\sqrt{2}}\left(\mu^{2}-1\right)\breve{J}^{2}_{0}{}^{\dagger}\left(x;i\right) \\ & - S^{2}_{0}{}^{\rm{old}}\left(x;i\right)
, \label{EQS20delta}
\end{split}\end{equation}
where $\bar{J}^{2}_{0}{}^{\dagger}$ and $\breve{J}^{2}_{0}{}^{\dagger}\left(x\right)$ result from the substitution of $J^{0}_{0}{}^{\dagger}\left(x\right)$ and $J^{2}_{0}{}^{\dagger}\left(x\right)$ into Eqs.\eqref{EQJ20bar} and \eqref{EQJ20conv}.

In summary, at each iterative step we solve the system of equations \eqref{EQMT11} in order to obtain the correction of the $J^{0}_{0}\left(x\right)$ radiation field. Then, we use this result to solve Eq. \eqref{EQMT7}, which gives us the correction for the $S^{0}_{0}$ source function. Finally, Eq. \eqref{EQS20delta} gives us the correction for the $S^{2}_{0}\left(x\right)$ source function.

\subsection{Convergence}\label{S33}

  The numerical method presented in the last section makes use of Jacobi's iterative method both for the line and the continuum part. The simpler alternative of this method is using Lambda iteration for the continuum, which converges provided that the continuum opacity is weak enough with respect to that of the line. The method presented can solve both the CRD line case without continuum opacity and the coherent continuum problem without line opacity, two problems that have different convergence rates.

  In order to illustrate this property of the numerical method we show the convergence rate for three different cases: i) CRD line for a $J_{u} = 1 \rightarrow J_{l} = 0$ transition without continuum, ii) Coherent continuum without line, iii) Non-coherent continuum without line.

\begin{figure}[htb]
\centering
\includegraphics[width=8.6cm]{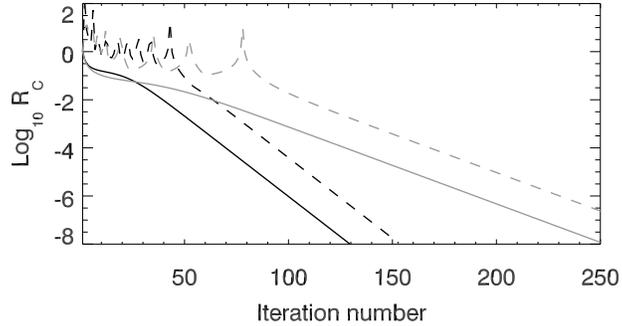}
 \caption{Maximum relative change of $S^{0}_{0}$ (solid lines) and $S^{2}_{0}$ (dashed lines) at each iterative step for a (CRD) resonance line without continuum (black lines), and for the continuum case  without line (gray lines). We point out that the convergence rates for the coherent and non-coherent cases are indistinguishable. For the continuum case the source function is frequency dependent, but here we take a fixed frequency because the convergence rate is virtually identical for all of them.}
\label{FIGRC}
\end{figure}

  For the first case, we take a Gaussian profile with $\Delta x = 0.1$ (distance between consecutive points in the frequency grid) and $\epsilon = 10^{-4}$. For the continuum cases, we take $s = 10^{-4}$ and, for the non-coherent case, $w = 11.7$ (width of the redistribution profile). We suppose an isothermal atmosphere and we solve with $\Delta z = 0.5$ (distance between consecutive points in the height grid, in units of the opacity scale height) and $N_{\mu} = 60$ Gaussian nodes for angular integration in each hemisphere. We present the corresponding convergence rates in Fig. \ref{FIGRC}. 

\begin{figure}[htb]
\centering
\includegraphics[width=8.6cm]{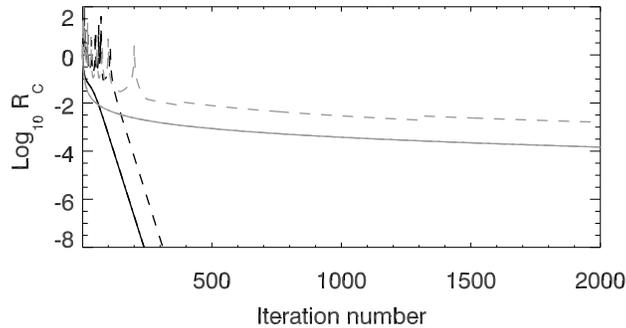}
 \caption{Maximum relative change of $S^{0}_{0}$ (solid lines), $S^{2}_{0}$ (dashed lines) and $J^{0}_{0}\left(x\right)$ (coincident with the solid line) at each iterative step for the CRD line transfer problem with continuum using the method described in section \ref{S32} (black lines) and using Lambda iteration for the continuum part (gray lines).}
\label{FIGRC2}
\end{figure}

  To demonstrate the virtue of the method with respect to the continuum treatment, we solve a problem where we include both a weak line and continuum, but using Lambda iteration for the continuum part. We take the same parameters used in Fig. \ref{FIGRC} and $\chi_{l}/\chi_{c} = 10$. This is a weak spectral line case, so the final rate of convergence is greatly influenced by the Lambda iteration of the continuum, that has a very poor convergence rate (see Fig. \ref{FIGRC2}).

\begin{figure}[htb]
\centering
\includegraphics[width=8.6cm]{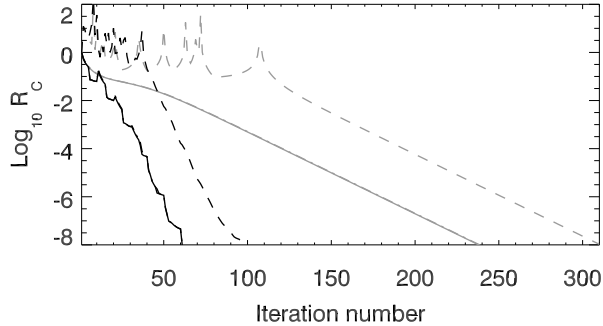}
 \caption{Maximum relative change of $S^{0}_{0}$ (solid lines), $S^{2}_{0}$ (dashed lines) and $J^{0}_{0}\left(x\right)$ (coincident with the solid line) at each iterative step for the CRD line transfer problem with coherent and non-coherent continuum, with (black lines) and without (gray lines) NG acceleration.}
\label{FIGRC3}
\end{figure}

  The code can also use \cite{NG1974} acceleration to decrease the total computing time. To show its efficiency we solve the problem of Fig. \ref{FIGRC2} using NG acceleration of third order. The number of iterative steps needed to reach convergence is greatly reduced without increasing significantly the computing time at each iterative step (see Fig. \ref{FIGRC3}). 

\subsection{Numerical Considerations}\label{S34}

  The precision of the numerical method depends on the parameters of the discretization in space, angles and frequencies. In the figures that are shown in Section \ref{S4} we take the following discretizations. A spatial height axis from $z_{\rm min} = -16$ to $z_{\rm max}=11$ or $16$ (this is more than needed to have an optically thick atmosphere at the bottom and an optically thin surface) with $\Delta z = 0.1$ or $0.5$, with the height $z$ measured in units of the opacity scale height. We use Gaussian quadrature with $60$ nodes at each hemisphere and a frequency axis that reaches $320$ Doppler widths with $\Delta x = 0.125$ in the core and with $\Delta x$ increasing with the distance to the resonance frequency ${\nu}_{0}$ until having $\Delta x = 15$ in the far wings.

   In order to demonstrate the reliability of our radiative transfer code, we solve the radiation transfer problem in a plane-parallel homogeneous atmosphere, relying on the fact that the $\sqrt{\epsilon}$-law (\citealt{AvrettHummer1965}; generalized to the polarized case by \citealt{Ivanov1990,LandiBommier1994}) provides an exact analytical result for the solution of this problem. We solve two of the problems of section \ref{S33}: coherent scattering in the continuum without line and resonance line without continuum. 
  
  In the far wings of the line, the spectrum can be considered frequency independent. The source function equations are thus simplified as
\begin{subequations}\label{EQScoh}\begin{align}
S_{I} = & S^{0}_{0}{}^{c} + \frac{1}{2\sqrt{2}}\left(3\mu^{2}-1\right)S^{2}_{0}{}^{c}
, \label{EQSIcoh} \\ 
S_{Q} = & \frac{3}{2 \sqrt{2}}\left(\mu^{2} - 1\right) S^{2}_{0}{}^{c}
, \label{EQSQcoh}
\end{align}\end{subequations}
with
\begin{subequations}\label{EQScoh2}\begin{align}
S^{0}_{0}{}^{c} = & s B_{\nu} + (1 - s)J_{0}^{0}
, \label{EQSIcoh2} \\ 
S^{2}_{0}{}^{c} = & (1 - s)J^{2}_{0}
. \label{EQSQcoh2}
\end{align}\end{subequations}
  The $\sqrt{\epsilon}$-law gives us the relation $\left(S^{0}_{0}{}^{c}\right)^{2} + \left(S^{2}_{0}{}^{c}\right)^{2} = s$. In Table \ref{TBLCOH} we show the relative error between the numerical result and this analytical relation; the agreement is very satisfactory. 

\begin{table}[htbp]
\centering
\caption{Verification of the $\sqrt{\epsilon}$-law for Different Values of $s$}
\begin{tabular}{ccc}
\hline \hline\\[-2.5ex]
$s$ & $\left(S_{0}^{0}{}^{c}\right)^{2} + \left(S^{2}_{0}{}^{c}\right)^{2} $ & Error (\%) \\[0.5ex]
\hline
$10^{-4}$ & $1.001\cdot 10^{-4}$   & $0.1$  \\
$10^{-3}$ & $1.0006\cdot 10^{-3}$ & $0.06$   \\
$0.01$      & $0.01002$                    & $0.02$  \\
$0.1$        & $0.100004$                  & $4\cdot 10^{-3}$  \\
$0.5$        & $0.500002$                  & $3\cdot 10^{-4}$  \\
$0.9$        & $0.90000007$              & $7\cdot 10^{-6}$  \\
\hline \hline
\end{tabular}
\label{TBLCOH}
\end{table}

   In the absense of continuum the source function equations become 
\begin{subequations}\label{EQSlinepure}\begin{align}
S_{I} =  &  S^{0}_{0}  + w^{\left(2\right)}_{J_{u} J_{\ell}}\frac{1}{2 \sqrt{2}}\left(3 \mu^{2} - 1\right) S^{2}_{0}
, \label{EQSIlinepure} \\
S_{Q} = & w^{\left(2\right)}_{J_{u} J_{\ell}}\frac{3}{2 \sqrt{2}}\left(\mu^{2} - 1\right) S^{2}_{0}
, \label{EQSQlinepure}
\end{align}\end{subequations}
with
\begin{subequations}\label{EQSlinepure2}\begin{align}
S^{0}_{0} = & \epsilon B_{\nu} + \left(1 - \epsilon\right)\bar{J}^{0}_{0}
, \label{EQSIlinepure2}\\
S^{2}_{0} = & w^{\left(2\right)}_{J_{u} J_{\ell}}\frac{1}{1 + \left(1 - \epsilon\right)\delta^{\left(2\right)}}\bar{J}^{2}_{0}
. \label{EQSQlinepure2}
\end{align}\end{subequations}
In Table \ref{TBLPUL} we check the $\sqrt{\epsilon}$-law for this line transfer problem with $\delta^{\left(2\right)} = 0$; the law is satisfied with good agreement.

\begin{table}[htbp]
\centering
\caption{Verification of the $\sqrt{\epsilon}$-law for Different Values of $\epsilon$.}
\begin{tabular}{ccc}
\hline \hline\\[-2.5ex]
$\epsilon$ & $\left(S_{0}^{0}\right)^{2} + \left(S^{2}_{0}\right)^{2}$ & Error (\%) \\[0.5ex]
\hline
$10^{-4}$ & $1.0005\cdot10^{-4}$ & $0.05$  \\
$10^{-3}$ & $1.0003\cdot10^{-3}$ & $0.03$   \\
$0.01$ & $0.010001$ & $0.01$  \\
$0.1$ & $0.100003$ & $3\cdot10^{-3}$  \\
$0.5$        & $0.500001$ & $3\cdot10^{-4}$  \\
$0.9$        & $0.90000006$ & $7\cdot10^{-6}$  \\
\hline \hline
\end{tabular}
\label{TBLPUL}
\end{table}

  In the next section we apply our radiative transfer code to some particular cases, where we have both line and continuum. We study some of the effects of the non-coherence of the scattering.

\section {Illustrative Examples}\label{S4}

   In this section we present some results of radiative transfer calculations in some model atmospheres. First, we make calculations in Milne-Eddington atmospheres with constant opacity ratios, because they are suitable for understanding the physics involved. Secondly, we suppose some ad-hoc variation with height of the properties of a model atmosphere with a temperature minimum and a chromospheric temperature rise. In both cases we consider line transitions with and without intrinsic polarization, the last case being quite interesting in terms of the emergent fractional polarization profile.

\subsection{Non-coherent Scattering in Milne-Eddington atmospheres}\label{S41}

   We study the interaction between a resonance line and the continuum radiation for two cases where non-coherent scattering in the continuum is taken into account or neglected. We assume a Milne-Eddington atmosphere with constant ratios between the different opacities involved. The important parameters in this model are the ratio between the opacity of the line and the continuum, $r =\chi_{l}/\chi_{c}$, and the relative weight of the thermal part to the total opacity of the continuum, $s = \kappa/\chi_{c}$. 

  We assume an intrinsically unpolarizable resonance line ($J_u = J_\ell =1/2$) and an intrinsically polarizable line ($J_u = 1$, $J_\ell = 0$). In both cases we solve the radiative transfer problem for a strong line $(r = 1000)$ and for a weak line $(r = 10)$ with $\epsilon = 10^{-4}$. For the continuum redistribution width we take $w = 11.7$ (value that we choose thinking in a forthcoming application to a realistic model; in particular, this number is the ratio between the Doppler widths of Barium and Hydrogen), and different values of $s$. We use a Milne-Eddington atmosphere with slope $3/2$.

\begin{figure*}[htp]
\centering 
\includegraphics[width=17.2cm]{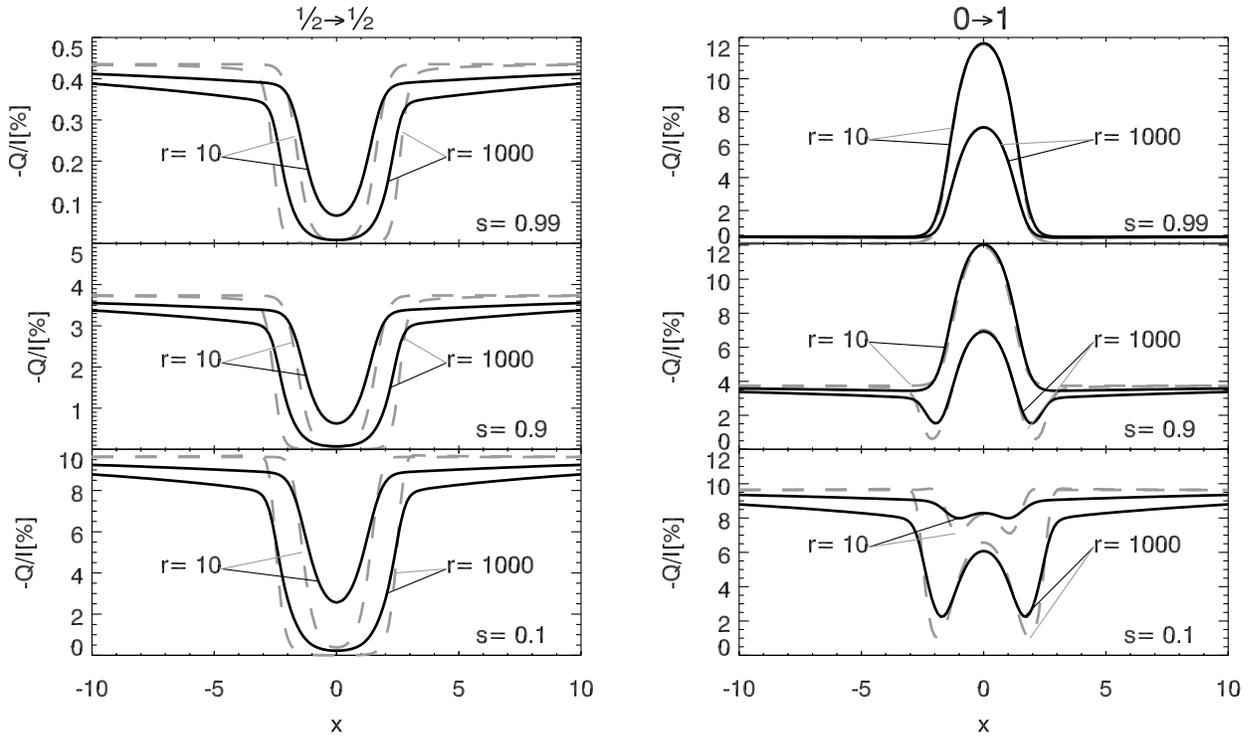}%
\caption{Emergent fractional polarization close to the limb $(\mu = 0.1)$ in a Milne-Eddington atmosphere for three different values of $s = \kappa/\left(\kappa+\sigma\right)$. The left panels are for an intrinsically unpolarizable transition ($J_u = J_\ell =1/2$) and the right panels are for a polarizable transition ($J_u = 1$, $J_\ell = 0$). Black solid lines correspond to the non-coherent scattering case, while gray dashed lines are for the case of coherent scattering in the continuum. The considered line strength values ($r$) are given in the panels.}
\label{FIGME}%
\end{figure*}

The non-coherent continuum scattering produces changes in the shape of the emergent fractional polarization profile, as has already been demonstrated in previous works (see Sect. \ref{S1}). When we study a strong ($r = 1000$) unpolarizable line, the coherent profile gives zero polarization in the core of the line, as expected. It is interesting to note that the redistribution produced by the non-coherent scattering polarizes the core of the line, although its $Q/I$ amplitude lies always below the continuum polarization level, i.e., the line always depolarizes the continuum. Thus, aside from being wider, the fractional polarization profile also shows non-zero polarization in the line core. The same happens to the $Q/I$ profile in the case of a weak ($r = 10$) intrinsically unpolarizable line (see Fig. \ref{FIGME}, left). The change in the $Q/I$ profile is larger for weaker lines and smaller $s$ values (or, equivalently, the more important is the scattering in the continuum).

   If we consider an intrinsically polarizable line, in order to obtain a noticeable change in the fractional polarization profile due to the non-coherent scattering in the continuum, we need the scattering coefficient $\sigma$ to be dominant over the thermal absorption term (small $s$). The smaller $s$, the more the polarization profiles changes. In all the cases shown in the right panels of Fig. \ref{FIGME} the intrinsic polarization of the line is dominant in its core and the non-coherence smoothes and broadens the fractional polarization profile in the wings of the line for small enough values of $s$. 

\subsection{Non-coherent Scattering in a Stratified Atmosphere with a Chromospheric Temperature Rise}\label{S42}

\begin{figure}[htp]
\centering 
\includegraphics[width=8.6cm]{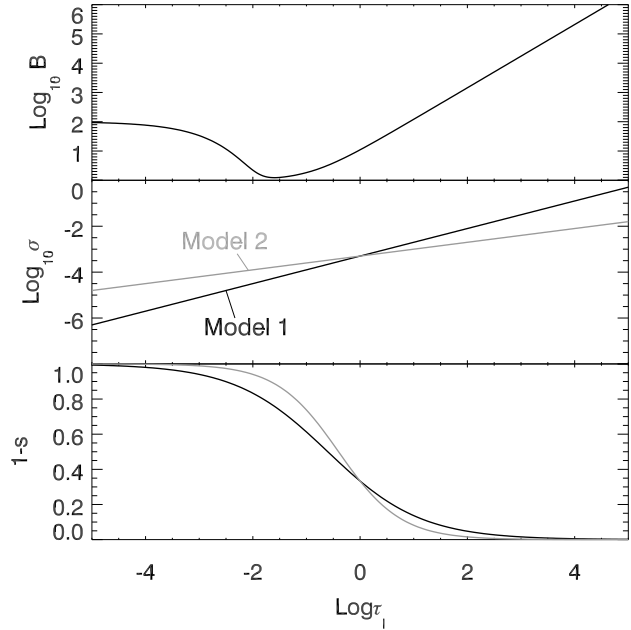} 
\caption{One dimensional atmospheric models. These two models have the same temperature, integrated line opacity $\chi_{l}$ and continuum thermal absorption $\kappa$ variations, but different behaviors for the continuum scattering coefficient $\sigma$. The upper panel shows the Planck function versus the integrated line optical depth. The middle panel shows the scattering coefficient $\sigma$ versus the optical depth, with the black line indicating the model $1$ and the gray line the model $2$. The bottom panel shows the quantity $1 - s = \sigma/\left(\kappa + \sigma\right)$, and we point out that it has the typical variation that can be found in semi-empirical models, such as those of \cite{FontenlaAvrettLoeser1993}.} 
\label{FIGVARMO}%
\end{figure}

    We assume now a certain height variation of the parameter $s$ and of the Planck function in order to obtain a more realistic stratification in the model atmosphere. Inspired by semi-empirical models of the solar atmosphere, we choose $\sigma$ in a way such that $1-s$ tends to unity near the surface and goes to zero at the bottom of the atmosphere (see Fig. \ref{FIGVARMO}, and note that $1 - s = \sigma/\left(\kappa + \sigma\right)$). We use two models that differ in the scattering coefficient. The variation with height of the scattering coefficient in the model $1$ is larger than in the model $2$ and the value of the scattering coefficient is the same at the height where the line integrated optical depth is unity. With these atmospheric models, we solve the two-level atom line transfer problem with $\epsilon = 10^{-4}$ and a non-coherent scattering redistribution width $w=11.7$, both for an intrinsically unpolarizable line ($J_u = J_\ell =1/2$) and for a polarizable one ($J_u = 1$, $J_\ell = 0$)

\begin{figure*}[htp]
\centering 
\subfigure{\includegraphics[width=8.6cm]{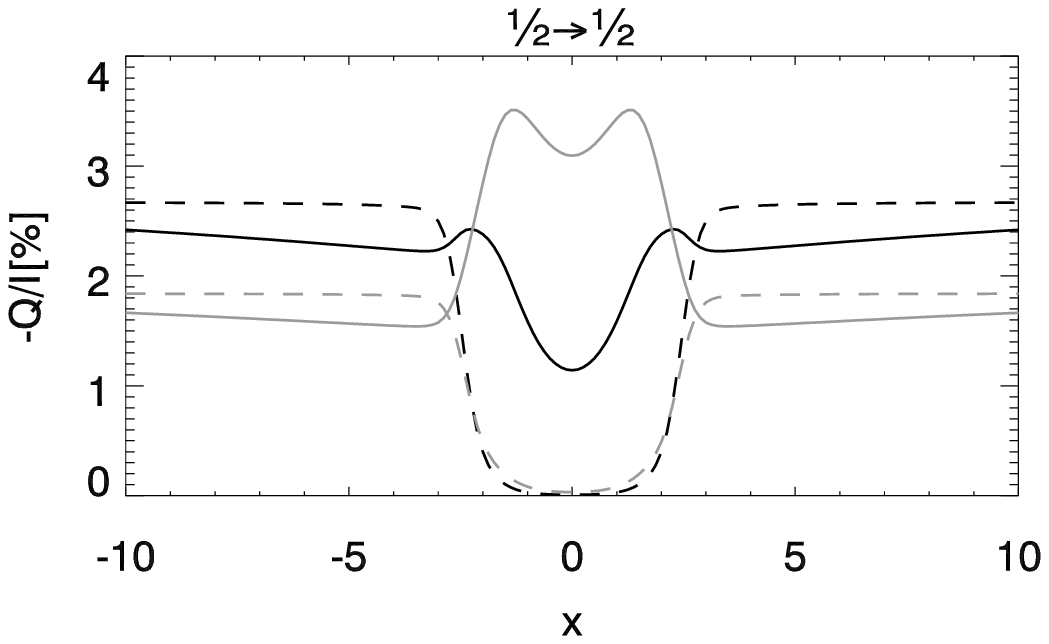}}%
\hspace{8pt}%
\subfigure{\includegraphics[width=8.6cm]{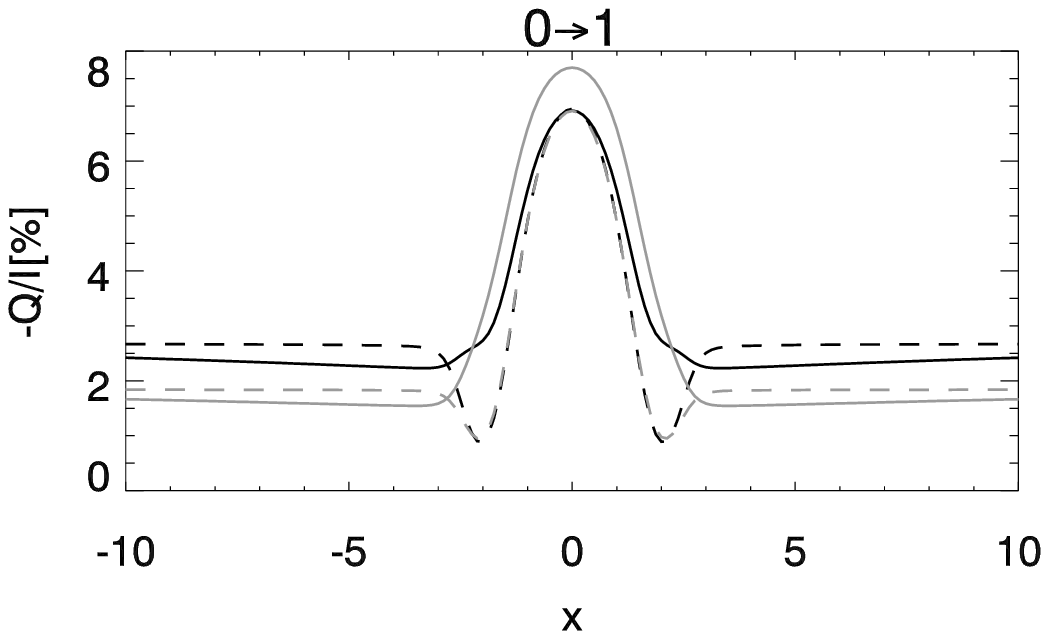}}
\caption{Emergent fractional polarization close to the limb ($\mu = 0.1$). The left panel shows an intrinsically unpolarizable transition ($1/2\rightarrow1/2$) and the right panel shows a polarizable one ($1\rightarrow0$). Solid lines show the non-coherent scattering solution, while the dashed lines the coherent scattering one. The black and gray lines correspond to the two different models indicated in Fig. \ref{FIGVARMO}} 
\label{FIGVARQI}%
\end{figure*}

   For a strong and unpolarizable line, the coherent profile shows a strong depolarization in the core (Fig. \ref{FIGVARQI}, dashed lines). However, the non-coherent $Q/I$ profile is strongly modified because the scattering redistribution produces polarization in the core of the lines (Fig. \ref{FIGVARQI}, solid lines). For the model $1$ (see Fig. \ref{FIGVARMO}), we see that the line does not fully depolarize the continuum, but the core is polarized and the profile is wider. For the model $2$ the non-coherent scattering generates an emission $Q/I$ profile (see Fig. \ref{FIGVARQI}).

  For a strong polarizable line, the coherent $Q/I$ profile shows the expected polarization emission in the core of the line. For the model $1$ the polarization in the core of the line does not change, and the main effect of the non-coherent scattering is the smoothing of the peaks in the wings of the line and the broadening of the $Q/I$ profile. For the model $2$, the redistribution is able to change even the polarization in the core of the line, while producing a smoother and wider $Q/I$ profile.

What we want to emphasize with Fig. \ref{FIGVARQI} is that \emph{the non-coherent scattering in the continuum can be important and, under certain conditions, there can be an emission feature in the fractional linear polarization profile even when a total depolarization is expected.}

  Finally, we study the influence of the mass of the scatterer. To this aim, we take a Milne-Eddington atmosphere with slope $3/2$ and the $\sigma$ variation of the model $2$ in Fig. \ref{FIGVARMO}. We solve the radiative transfer problem for a $J_{\ell} = J_{u} = 1/2$ transition with gaussian absorption profile, with different widths of the redistribution function (this width is inversely proportional to the square root of the mass of the scatterer).

\begin{figure}[htb]
\centering 
\includegraphics[width=8.6cm]{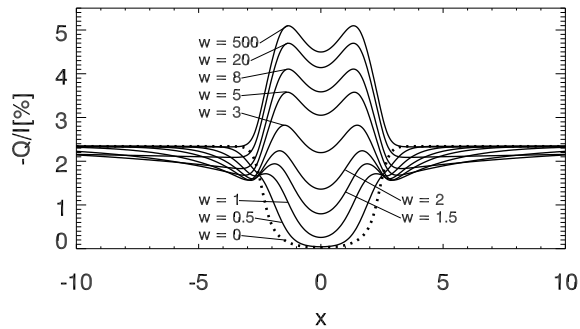} 
\caption{Fractional polarization profiles close to the limb ($\mu = 0.1$) for a $1/2\rightarrow 1/2$ transition in a Milne-Eddington atmosphere, with the $\sigma$ and $1 - s$ variations given by the model $2$ of Fig. \ref{FIGVARMO}, for different widths of the scattering redistribution function. The dotted line shows the coherent case.} 
\label{FIGVARWpro}%
\end{figure}

\begin{figure}[htb]
\centering 
\includegraphics[width=8.6cm]{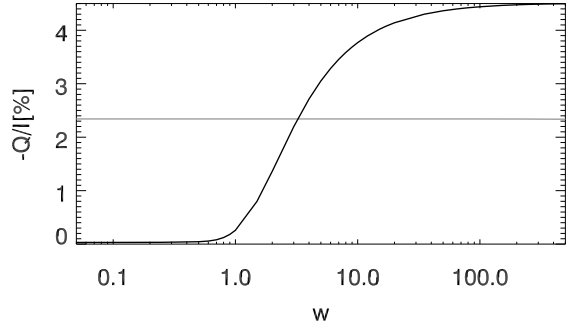} 
\caption{Line center fractional linear polarization close to the limb ($\mu = 0.1$) for a $1/2\rightarrow 1/2$ transition in a Milne-Eddington atmosphere (with $1-s$ given by the model $2$ of Fig. \ref{FIGVARMO}) versus the width of the scattering redistribution function. The gray line represents the fractional continuum polarization amplitude.} 
\label{FIGVARW}%
\end{figure}

   For small widths we approach the coherent case, where the line is depolarized. As we increase the width of the velocity redistribution profile, the linear polarization in the core of the line increases. In Fig. \ref{FIGVARWpro} we show some fractional polarization profiles for several values of the widths of the redistribution profile. If we take the center of the line as reference and we plot the fractional polarization at this frequency versus the widths of the redistribution profile, we obtain the curve shown in Fig. \ref{FIGVARW}, where we have also indicated the continuum fractional polarization level. In this figure we can see that from a given value of $w$ the line-center signal of the $Q/I$ profile lies above the continuum level and increases to an asymptotic value. From this figure we can infer that, for a given value of $s$, Thomson scattering  (whose associated width is approximately $43$ times the Doppler width of Hydrogen) produces a greater polarization than Rayleigh scattering for an intrinsically unpolarizable line.
   
\section{Conclusions}

In this paper we have studied the radiative transfer problem of resonance line polarization taking into account non-coherent continuum scattering, paying particular attention to the fractional linear polarization $Q/I$ signals that can be produced around the core of intrinsically unpolarizable lines. We used the two-level atom model with CRD and angle-averaged non-coherent scattering in the continuum. To numerically solve this type of radiative transfer problem we developed a Jacobian iterative method for the line and continuum source functions, which yields a fast convergence rate even in the case of very small line strengths. The formulation of the numerical method makes it very suitable for a direct generalization to partial frequency redistribution and angle-dependent non-coherent scattering.

We have shown that, under certain conditions, the non-coherent continuum scattering can change dramatically the core spectral region of the emergent $Q/I$ profile with respect to that calculated assuming coherent continuum scattering. Interestingly, $Q/I$ polarization signals above the continuum level can be generated in the core of intrinsically unpolarizable $(1/2\rightarrow 1/2)$ lines (i.e., in spectral lines that were expected to simply depolarize the continuum polarization level). This result is of great potential interest for a better understanding of some enigmatic spectral lines of the second solar spectrum, which showed $Q/I$ line-center signals above the continuum polarization level in spite of resulting from transitions between levels that were thought to be intrinsically unpolarizable (see Stenflo et al. 2000). Of particular interest for a first application is the D$_1$ line of Ba {\sc ii} at 4934 \AA, especially because 82\% of the barium isotopes have nuclear spin $I=0$ (i.e., their D$_1$ line transition is indeed between an upper and lower level with total angular momentum $J_{u} = J_{\ell} = 1/2$). In fact, our preliminary calculations for the Ba {\sc ii} D$_1$ line (neglecting the contribution of the 18\% of barium that has hyperfine structure) suggest that under certain stellar atmospheric conditions the physical mechanism discussed in this paper can produce significant $Q/I$ emission features. 

Finally, we point out that the core of strong lines with intrinsic polarization are practically not affected by the non-coherent scattering. Therefore, the effects of the non-coherent scattering in the continuum are not always relevant and depend on the spectral line under study.

\acknowledgements
Financial support by the Spanish Ministry of Economy and Competitiveness through projects \mbox{AYA2010--18029} (Solar Magnetism and Astrophysical Spectropolarimetry) and CONSOLIDER INGENIO CSD2009-00038 (Molecular Astrophysics: The Herschel and Alma Era) is gratefully acknowledged.

\appendix 

  The $\Lambda^{K}_{Q}\left(x;i,j\right)$ operators and the $T^{K}_{0}\left(x;i\right)$ quantities that appear in Eqs. \ref{EQJxnum} are given by:
\begin{subequations}\label{EQLAM}\begin{align}
\Lambda^{0}_{0}\left(x;i,j\right) = & \frac{1}{2}\int_{-1}^{1}d\mu \, \Lambda\left(x,\mu;i,j\right)
, \label{EQLAM00} \\
\Lambda^{0}_{2}\left(x;i,j\right) = & \frac{1}{4\sqrt{2}}\int_{-1}^{1}d\mu \left(3\mu^{2} - 1\right)\Lambda\left(x,\mu;i,j\right)
,\label{EQLAM02} \\
\Lambda^{2}_{0}\left(x;i,j\right) = & \Lambda^{0}_{2}\left(x;i,j\right)
,\label{EQLAM20} \\ \begin{split}
\Lambda^{2}_{2}\left(x;i,j\right) = & \frac{1}{16}\int_{-1}^{1}d\mu\bigg[\left(3\mu^{2}-1\right)^{2} + 9\left(\mu^{2}-1\right)\bigg]\Lambda\left(x,\mu;i,j\right)
,\label{EQLAM22} \end{split} \\
T^{0}_{0}\left(x;i\right) = & \frac{1}{2}\int_{-1}^{1}d\mu \, T_{I}\left(x,\mu;i\right)
,\label{EQT00} \\ \begin{split}
T^{2}_{0}\left(x;i\right) = & \frac{1}{4\sqrt{2}}\int_{-1}^{1}d\mu\bigg[\left(3\mu^{2} - 1\right)T_{I}\left(x,\mu;i\right) + 3\left(\mu^{2} - 1\right)T_{Q}\left(x,\mu;i\right)\bigg]
.\label{EQT20} \end{split}
\end{align}\end{subequations}

\bibliographystyle{apj}

\end{document}